\documentclass[]{article}
\usepackage{amsmath}
\usepackage{amssymb}
\usepackage{amsfonts}
\usepackage{url}
\usepackage[affil-it]{authblk}
\usepackage{xfrac}
\newcommand{\ket}[1]{\left| #1 \right>} 

\begin{document}
\sloppy

\title{Quantum computing for pattern classification}

\author{Maria Schuld$^{a}$\footnote{Corresponding author: schuld@ukzn.ac.za }, Ilya Sinayskiy$^{a,b}$ and Francesco Petruccione$^{a,b}$}
\affil{$^{a}${\em{Quantum Research Group, School of Chemistry and Physics,  University of KwaZulu-Natal, Durban, KwaZulu-Natal, 4001, South Africa}}\\
$^{b}${\em{National Institute for Theoretical Physics (NITheP), KwaZulu-Natal, 4001, South Africa}}}\vspace{6pt}
\date{\today}

\maketitle

 It is well known that for certain tasks, quantum computing outperforms classical computing. A growing number of contributions try to use this advantage in order to improve or extend classical machine learning algorithms by methods of quantum information theory. This paper gives a brief introduction into quantum machine learning using the example of pattern classification. We introduce a quantum pattern classification algorithm that draws on Trugenberger's proposal for measuring the Hamming distance on a quantum computer  (CA Trugenberger, \emph{Phys Rev Let} 87, 2001) and discuss its advantages using handwritten digit recognition as from the MNIST database. \\


\vspace{1cm}

\section{Quantum methods for machine learning}
With the rapid growth in the volume of data that is transferred, stored and processed on a daily basis, innovative methods of machine learning become more and more important. Supervised machine learning algorithms infer an input-output relation from large sets of training data that consist of `correct' examples of mappings. In other words, the computer \textit{learns} from experience how to treat new inputs. A prominent example where a mapping needs to be learned is the pattern classification problem, in which a new data vector has to be assigned to one of a number of classes, given a set of correctly classified data vectors. A data vector thereby contains information on the features of the entity that is to be classified (for example the clicking behaviour of an online user, the structure of a molecule or the pixel of an image), and is also called \textit{feature vector}. Pattern classification is the abstract formulation of the problem of interpreting information, and it finds application in areas as diverse as information technology, the food industry or the financial sector. These tasks come to humans much more natural than to machines (e.g., when we recognise other humans as a response to the large amount of photons that enter our eyes in every second), and as a subdiscipline of artificial intelligence, machine learning is indeed inspired by the way of how our brain deals with data.\\

Quantum computing is a relatively new branch combining computer science and physics, in which the properties of small particles formulated in quantum theory are exploited to process information. Controlling quantum objects that encode information (so called qubits or qudits) is a highly nontrivial task, and the realisation of a mature quantum computer is still far from being accomplished. However, there is no lack of theoretical studies on the scope and power of quantum information. As part of these efforts, quantum information scientists recently realised that quantum computing could improve classical machine learning algorithms in three basic ways. First, subroutines that are costly on a classical computer when subjected to big data - such as the evaluation of an inner product or searching for a minimum distance -  could be executed on a quantum computer with a linear or even exponential speedup in complexity due to quantum parallelism \cite{lloyd13,rebentrost13,wiebe14}.  Second, from the perspective of quantum computing, quantum machine learning (from here on QML) opens up new possibilities for quantum information processing, such as quantum state classification \cite{sentis12,gutja10}. Third, especially in the area of intelligent agents and reinforcement learning, quantum physics offers unique types of logics that is often compared to fuzzy logics \cite{rigatos07,busemeyer12}.  \\

The advantage of computing with quantum objects is that data can theoretically be represented exponentially more compact in a so called quantum superposition of both the $0$ and $1$ state. On the downside, information retrieval is limited by the laws of measurement of a quantum system, which is a destructive process that changes it substantially preventing us from assessing all information at once \cite{sasaki01}. (However, as we will see later the probabilistic nature of the outputs to quantum algorithms can be valuable for pattern classification). The last decade of quantum information research provided a `toolbox' of algorithms that can be implemented on a potential quantum computer, which are the building blocks used to tackle the more sophisticated problems of QML. Here, we will add to these methods and propose a quantum pattern classification algorithm for binary feature vectors, which follows the principle of a distance weighted $k$-nearest neighbour method \cite{dudani76}. Our idea uses a variation of Trugenberger's  \cite{trugenberger01} subroutine to determine the Hamming distance  between two binary patterns on a quantum computer. \\

This paper is organised as follows. We will first give a brief introduction to quantum computing which can be skipped by readers familiar with quantum information theory. We then outline the problem of pattern recognition (Section \ref{pc}). In Section \ref{qpc} we briefly introduce into distance-based methods of pattern classification such as \textit{$k$-nearest neighbours}, translate the problem into the language of quantum physics and give an example of a quantum classification algorithm. We discuss its merit using handwritten digit recognition and give a general outlook in the context of quantum machine learning.

\section{Computing with quantum objects} \label{qc}

Quantum computing analyses the manipulation of quantum objects in order to solve computational problems\footnote{For a comprehensive introduction to quantum computing, see \cite{nielsen10}.}. A `quantum object' thereby refers to any particle or system of particles for which Newton's mechanics proves to be an insufficient description while quantum theory explains our observations. This is becomes important for the description of microscopic particles such as atoms, electrons or photons, and allows for entirely new ways of information processing on a microscopic scale.\\

The equivalents to bits on a classical computer are quantum objects with two distinct configurations or \textit{states}, called qubits\footnote{Note that \textit{qudits} would be the generalisation to $d$-dimensional states.}, which can have various physical implementations such as the energy of atoms or the polarisation of photons. But if bits are either carrying a signal encoding a $0$ or $1$, qubits use the superposition principle of quantum objects to be in both states `at the same time'. In the notation for quantum states, this looks like 
\[\ket{\psi} = \alpha \ket{0} + \beta \ket{1} , \qquad |\alpha|^2 +|\beta|^2 = 1,\]
where $\alpha, \beta$ are complex numbers called \textit{amplitudes} and $\ket{\cdot}$ represents a state vector describing a quantum object. Later on the \textit{phase} $\phi$ of a qubit becomes important, which is a part of the amplitude $\alpha=\tilde{\alpha} e^{i\phi}$. Quantum theory is built around the observation that the squared amplitudes $ |\alpha|^2 ,|\beta|^2 $ denote the probability to measure the qubit either  in state $\ket{0} $ or $\ket{1} $. A qubit state is thus not characterised by whether it is in the `$0$' or `$1$' state, but by \textit{how likely it is to measure it in either of them}. Computations can work on both states at the same time, a fact that is often referred to as quantum parallelism. \\

The power of quantum information processing becomes apparent if we consider a system of $n$ qubits each with the two available states $\{\ket{0}, \ket{1}\}$. The quantum system can be put into a superposition of all $2^n$ combinations $\{\ket{00...00},\ket{00...01},...,\ket{11...11}\}$ and an algorithm can work on all these configurations in parallel. However, quantum computing is always constrained by the probabilistic nature of the results, as well as the destruction caused by measurement. After a qubit has been measured to be either $\ket{0} $ or $\ket{1} $, the state `collapses' into the measurement result and will subsequently only produce the same output. We can therefore only access a limited amount of information from the result, and the output is of probabilistic nature (i.e. evolving and measuring the same system several times produces a distribution of results, of which the most likely result can be regarded as the output of the computation). This is why it is rather difficult to come up with powerful algorithms for a quantum computer \cite{nielsen10}.\\

It is  important for the following to introduce some formal basics of quantum information theory, but the interested reader shall be referred to  \cite{nielsen10}. The discrete states of a quantum object (such as the above mentioned polarisation or energy level) are mathematically modelled as vectors in a $d$-dimensional Hilbert space $\mathcal{H}_d$. For qubits, $d$ equals $2$, and a system of $n$ qubits that encodes a binary string of the same length can be described by vectors in $\mathcal{H}_2 \otimes ...\otimes \mathcal{H}_2$ (remember that the $d$-dimensional generalisation of a qubit is then called a `qudit'). Transformations from one vector to another that obey the general laws of quantum theory are represented by unitary operators $U$ with the property $U^{\dagger}U = 1$ where $U^{\dagger}$ is the hermitian conjugate. These unitary transformations define the dynamics of the quantum system and quantum algorithms can be represented by a sequence of such operations on an input quantum state. \\

In quantum computation, these unitary transformations are called `quantum gates', since they correspond to classical gates that manipulate bits. Some standard $1$-qubit gates are the $X$-gate that flips the state of a qubit, the $Z$-gate that changes the sign of its amplitude, or the Hadamard or $H$-gate that creates a superposition $\frac{1}{\sqrt{2}} (\ket{0} \pm  \ket{1})$ from $\ket{0}$ ($+$) or $\ket{1}$ ($-$) respectively. A central $2$-qubit gate is the controlled-$\mathrm{NOT}$ operation $\mathrm{cNOT}$ which only flips the state of a second qubit if the first one is in state $\ket{1}$. The $\mathrm{cNOT}$ together with standard single qubit gates form in fact a universal set for quantum computation \cite{nielsen10}. A more general formulation for a quantum gate that is derived from fundamental quantum theory based on the Schr\"odinger equation can be described by a unitary transformation $U = e^{-i\mathrm{H}t}$ where $\mathrm{H}$ denotes a hermitian operator called Hamiltonian.

\section{The $k$-nearest neighbour algorithm for pattern classification}\label{pc}

In a pattern classification problem we want to assign one out of a number of classes to a pattern, according to a rule learned from a set of example classifications. It is thus a problem of \textit{supervised} learning, or learning from training data. This abstract formulation contains an impressive range of important decision problems in real life. For example, a doctor diagnosing a disease given a number of symptoms and his experience from other cases, an email being automatically marked as `important' or `spam' on the grounds of previous emails, or a handwritten digit on a postal envelope being recognised by a scanning device. Even more, our human thinking process can be described through decision problems, for example when we `recognise' (= classify) people, things and smells around us, or when we classify a situation as dangerous or not depending on sensual stimuli. Some authors replace the term pattern classification by \textit{pattern recognition}, which is a more generalised expression as it also looks at the problem of seeing patterns without classifying it, as well as \textit{template matching} or \textit{associative memory}, in which a close example from learnt data is retrieved upon an input.  \\

Describing the pattern classification problem more precisely, we are given a set of $n$-dimensional data vectors $\vec{v}^k$ and their class assignments $c^p$, $\mathcal{T}=\{ (\vec{v}^{p}, c^p) \}_{p=1,..., N}$ that makes up the `training set' to our problem. Each of the vectors encodes $n$ features $v_{i}^p$. These features may represent the grey shade of a certain pixel, information on whether a patient has had cancer in his or her family, or the number of times a certain word occurs in a text sample. The features are given as by binary, integer or real-valued numbers, while the class $c^p$ of a feature vector is often encoded by a finite number $d$ of positive integers $c\in\{1,...,d\}$. Also given is an unclassified input vector $\vec{x}$ from the same vector space as the training vectors, encoding $n$ features. The task of pattern classification is to match the new vector $\vec{x}$ to a class, using information from the training data. This is usually done by defining some distance measure and assigning the new input vector to the class whose members are the most `similar' in terms of this distance. A common distance measure is the Euclidean metric or in case of binary features, the Hamming distance \cite{alpaydin04} (the number of differing bits on two binary strings  \cite{hamming50}).\\

\begin{figure}[t]
  \centering    \includegraphics[width=0.4\textwidth]{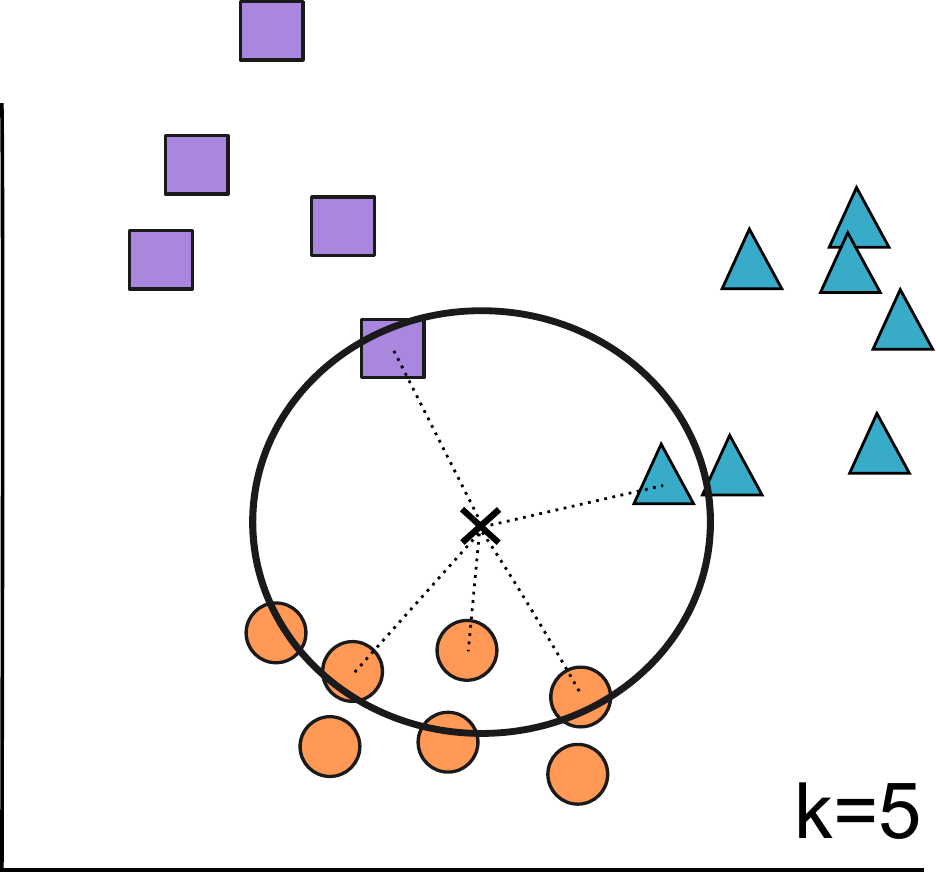} \caption{Illustration of the kNN method of pattern classification. The new vector (black cross) gets assigned to the class that the majority of its $k$ closest neighbours have (in this case it would be the orange circle shape). In this example, $k$ is set to 5.} \label{figure1}
\end{figure}

The discipline of machine learning developed a number of algorithms to solve the problem of pattern classification. One the most famous is the $k$-nearest neighbour (kNN) method \cite{harrington12,rogers12}. Given a training set $\mathcal{T}$ stored in a memory, the $k$ training vectors closest to the input vector are selected. The class to which the majority of these neighbours belong consequently gets assigned to the input vector (see Fig. \ref{figure1}).  
There are many variations to this simple method. For example, in the distance-weighted kNN, the neighbours get weighted by their distance to the input vector, so that closer neighbours make a bigger contribution to which class gets selected \cite{dudani76}. Another variation includes to preprocess the training data and calculate the centroid $\vec{\bar{v}}_{c} = |\vec{x}- \frac{1}{L}\sum_{l \in c} \vec{v}^l|$ of each class $c \in \{1,...,d\}$ ($l = 1,...,L_c$ is the index for the members of class $c$). The input vector then becomes part of the class with the closest centroid vector.\\ 

The advantage of the kNN method and its variations is not only their simplicity. They are \textit{nonparametric} examples of supervised learning, since they do not require initial information on the distribution of vectors \cite{duda12}. The only assumption on the data used is that \textit{similar inputs have similar outputs} \cite{alpaydin04} (in our case, similar input vectors should be in general members of the same class). An important task is to choose an optimal parameter $k$, and in the original kNN method, a balance between noise reduction and maintaining the locality information has to be found (for example, for $k\rightarrow \mathrm{all}$, then the class assignment would always result in the class with the most members). The distance-weighted kNN version has the advantage of being independent of the choice of $k$, as the ``number of nearest neighbors is implicitly hidden in the weights'' \cite{hechenbichler04}. The quantum algorithm introduced in the following is based on the same principle as kNN, namely assuming that `close' feature vectors carry the class that is to be assigned to the new vector.

\section{Quantum pattern classification}\label{qpc}

Translating the pattern classification problem into the language of quantum physics reads as follows. Our feature data set is represented by quantum states $\{\ket{v_1^p...v_n^p, c^p} \}\in \mathcal{H}_{2}^{\otimes n} \otimes \mathcal{H}_d$ where $p = 1,...,N$ runs over the training states and the class of feature vector $\vec{v}^p$ is stored in the qudit $\ket{c^p}$. The product state of $n$ $2$-dimensional Hilbert spaces thereby represents the feature space, while the additional qudit in $\mathcal{H}_d$ encodes the $d$ possible classifications. The input vector is a quantum state $\ket{x_1,...,x_n}$ from the feature Hilbert space $\mathcal{H}_{2}^{\otimes n}$. As we are now dealing with quantum information, classical data either has to be translated into quantum states, or -as suggested in \cite{giovannetti08,lloyd13}- taken from some form of a quantum random access memory (especially if the machine learning algorithm is a subroutine to a larger computation on a future quantum computer). 
\subsection{Related work}

Many of the textbook machine learning methods already faced attempts to be translated into quantum physics (for a detailed review, see \cite{schuld14c}). Amongst them are support vector machines \cite{rebentrost13}, decision trees \cite{lu14}, principal component analysis \cite{lloyd13b}, learning from membership queries \cite{servedio04}, neural networks \cite{ventura00,zhou08} and clustering \cite{aimeur07,horn02,aimeur06}. Most contributions are dedicated to pattern recognition or classification tasks \cite{lloyd13,rebentrost13,gambs08,neven09,sasaki01,sentis12,wiebe14,schuetzhold02,trugenberger02}. Some of these proposals are based on the idea of taking a computationally expensive subroutine from an original machine learning algorithm and executing it more efficiently on a quantum computer \cite{lloyd13,wiebe14,schuetzhold02,rebentrost13}. In \cite{sasaki01,sentis12} we find an attempt to use the insights of Bayesian decision theory for the classification of unknown quantum states. Some use adiabatic quantum computing to solve a learning optimisation problem \cite{neven09,lloyd13}. A number of contributions also try to execute classical distance measures through quantum computation \cite{lloyd13,rebentrost13,wiebe14,trugenberger02}. Finally, some authors emphasize the observation that the theory of open quantum systems is close to machine learning methods based on Markov models \cite{clark14,barry14}. Despite this growing number of contributions, quantum machine learning is still a premature discipline, which derives its relevance from its potential to extend machine learning by a new paradigm, rather than from a given theoretical foundation. Although touched upon in several articles \cite{sasaki02,hunziker03}, there is yet no fundamental theory of how quantum information can in general be exploited for intelligent forms of computing. The expression `quantum learning' \cite{bonner03,sasaki02,gutja10}  is so far used interchangeably with the term `quantum machine learning' and simply refers to the various ideas brought forward in order to integrate quantum information into methods of machine learning or vice versa.  

\subsection{A quantum pattern classification algorithm}

\begin{figure}[t]
  \centering    \includegraphics[width=0.4\textwidth]{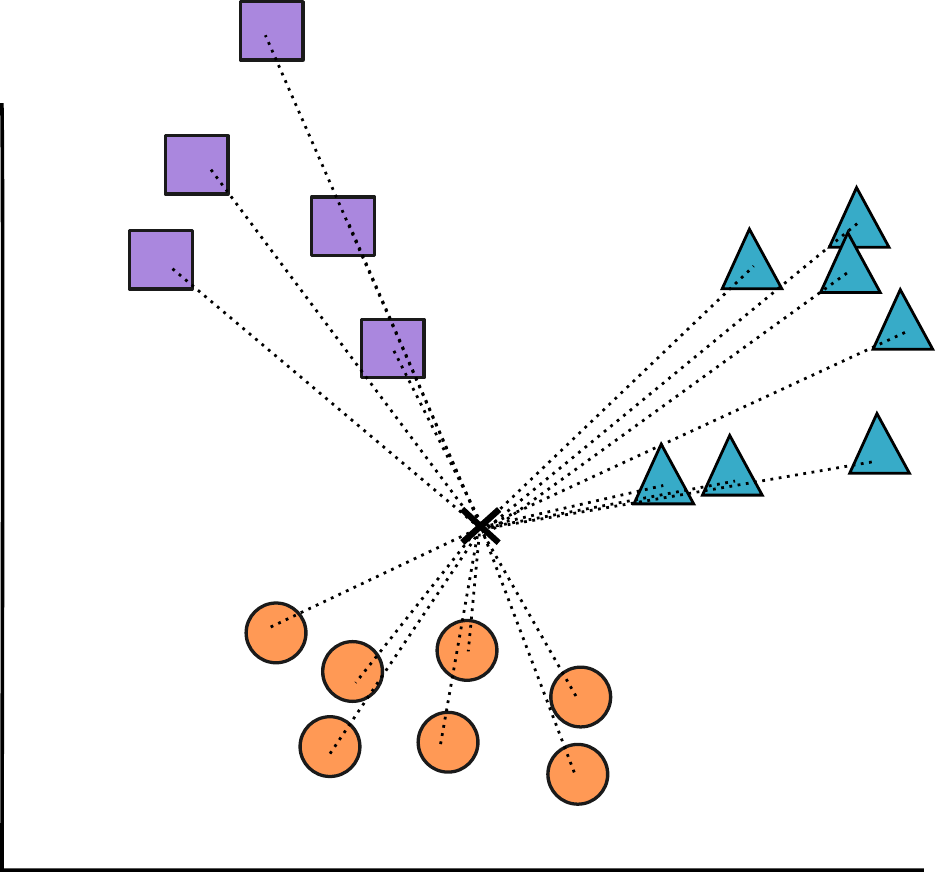} \caption{Illustration of the principle on which the quantum pattern classification is based. The new vector (black cross) gets assigned to the class with the closest members. As in Figure \ref{figure1}, this would be the class of orange circles.} \label{figure2}
\end{figure}

The quantum pattern classification (QPC) algorithm we present here uses the same distance-based classification principle as kNN, only that instead of chosing nearest neighbours, the distance of the entire training vector set is considered (see Figure \ref{figure2}). It draws on an algorithm presented in the context of quantum associative memory in \cite{trugenberger01}. The idea is to create a superposition of the training data set and `write' the Hamming distance to the input state into the amplitude of each vector in the superposition. Measuring the class-qudit then retrieves the desired class with the highest probability. Even more, if repeated enough times to achieve statistical significance, the algorithm leads to a probability distribution containing information on how close each class members are to the input vector. Note that the following requires an understanding of the circuit model of quantum computing that was touched upon in Section \ref{qc}, and readers not sufficiently familiar with quantum information theory might prefer to only consider the result in Eq (\ref{result}) and the discussion thereafter.\\

The initial step of the algorithm is to construct a `training set superposition' containing the training data,
\[\ket{T} = \frac{1}{\sqrt{N}} \sum_p \ket{v_1^p...v_n^p, c^p}. \]
While this `training phase' is trivial in the classical case, the efficient preparation of a quantum system in an arbitrary initial state is still an open problem, and also questions of quantum memory devices have not been resolved yet. However, algorithms to construct the initial state from a ground state can be found in \cite{trugenberger01,ventura00} and have linear complexity, just as accessing each bit from a classical memory would have. From this we construct the initial state
\[\ket{\psi_0} = \frac{1}{\sqrt{N}} \sum_p \ket{x_1...x_n;v_1^p...v_n^p, c^p ; 0}.  \]
It is made of three registers, the first containing the input state, the second  containing the superposition $\ket{T}$ and the third containing an ancilla qubit set to zero. In the first step, the ancilla is put into a superposition through a Hadamard gate, leading to
\[\ket{\psi_1} = \frac{1}{\sqrt{N}} \sum_i \ket{x_1...x_n;v_1^p...v_n^p, c^p } \otimes \frac{1}{\sqrt{2}} (\ket{0} + \ket{1}).  \]
For reasons of readibility we factor the ancilla state out for now. Following \cite{trugenberger01}, in a second step the Hamming distance between each qubit of the first and second register, 
\[d_k^{i} = \left\{   \begin{array}{l l}
		    1, & \quad \mathrm{if} \; \ket{v_k^p} = \ket{x_k},\\
   		    0, & \quad \mathrm{else,} 
  		\end{array} \right . \]
replaces the qubits in the second register. This is done by applying an $\mathrm{cNOT}(a,b)$-gate (see Section \ref{qc}) which overwrites the second entry $b$ with $0$ if $a=b$ and else with $1$. We use the $X$ gate to reverse the states in the second register, since in the end we want a strong `signal' for small Hamming distances. Note that the gates have no effect on the class and ancilla states. The second step consequently reads 
\begin{align*}\ket{\psi_2} &= \prod_k \mathrm{X}(x_k) \; \mathrm{cNOT}(x_k, v^p_k) \; \ket{\psi_1}  \\
&=\frac{1}{\sqrt{N}}\sum_p \ket{x_1...x_n;d_1^p...d_n^p, c^p} \otimes \frac{1}{\sqrt{2}}(\ket{0} + \ket{1})
.  \end{align*}
We then use the unitary operator
\[U = e^{ -i\frac{\pi }{2n} H}  \qquad\; H = 1 \otimes \sum_k \left(\frac{\sigma_z +1}{2}\right)_{d_k}  \otimes 1 \otimes (\sigma_z)_c\; , \]
to sum up the reverse single-qubit Hamming distances $d^p_k$ of each training vector $\ket{v_1^p...v_n^p}$ in order to write the total reverse Hamming distance between input vector and the $p$th training vector, $\bar{d}_H(\vec{x},\vec{v}^p)$,  into the phase of the $i$th state of the superposition (together with a negative sign if the ancilla qubit is $\ket{1}$). The state after the third step is consequently given by
\begin{multline}\ket{\psi_3} = U \ket{\psi_2} = \frac{1}{\sqrt{2N}} \sum_p  e^{ i\frac{\pi }{2n} \bar{d}_H(\vec{x},\vec{v}^p)}  \ket{x_1...x_n;d_1^p...d_n^p, c^p; 0} \\
+ e^{- i\frac{\pi }{2n} \bar{d}_H(\vec{x},\vec{v}^p)} \ket{x_1...x_n;d_1^p...d_n^p, c^p; 1} \label{btw}
.  \end{multline}
Another Hadamard transformation on the ancilla state, $H = 1 \otimes 1 \otimes 1 \otimes H_a$,  writes the phase information into the amplitudes,

\begin{multline*}\ket{\psi_4} = H \ket{\psi_3}
=\frac{1}{\sqrt{N}}\sum_p  \mathrm{cos}\left[ \frac{\pi}{2n} \bar{d}_H(\vec{x},\vec{v}^p) \right]  \ket{x_1...x_n;d_1^p...d_n^p, c^p; 0} \\
+ \mathrm{sin}\left[ \frac{\pi}{2n} \bar{d}_H(\vec{x},\vec{v}^p)\right]  \ket{x_1...x_n;d_1^p...d_n^p, c^p; 1} 
.  \end{multline*}

The ancilla does not only allow for this trick, but also gives us a possibility to test if the Hamming distance between the input we aim to classify, $\ket{x}$, and the states $\ket{v^p}, \; p = 1,...,P$ is on average large or small. If the new input is far away from most training patterns, we have a much higher probability to measure the ancilla in the state $\ket{1}$, if the input is close to many patterns we end up in state $\ket{0}$. Trugenberger in his quantum associative memory only accepts inputs that have a sufficiently high probability of an ancilla state in $\ket{0}$, arguing that only in this case an associative memory can be reliable. Although our QPC algorithm should not rely on the average distance between the input and the training vectors, for the following retrieval step we have to measure the ancilla until we get a $\ket{0}$ in order to retrieve the cosine part of the sum. Obviously, the closer the input is to the training set, the more likely that we succeed. Our simulations show that the probability for this measurement,

\[P(0_a) = \frac{1}{N}\sum_p  \mathrm{cos}^2\left[ \frac{\pi}{2n}\bar{d}_H(\vec{x},\vec{v}^p) \right], \]
is higher than $\frac{2}{3}$ for standard examples like the MNIST\footnote{The \textit{Mixed National Institute of Standards and Technology} database is a collection of handwritten digits and can be retrieved from \url{http://yann.lecun.com/exdb/mnist/} [last visit 19/9/2014]. } handwritten digit database. \\

There are two versions of how to proceed, one that corresponds to a ``$k \rightarrow \mathrm{all}$'' method assigning the class of vectors that are on average closer to the input, and another version that measures the pattern register and retrieves neighbours with a probability weighed by their distance, and chooses the class most represented by this pool.\\
 
Following the first version, the last step is a measurement of the class-qudit along the standard basis. This step varies from \cite{trugenberger01}, in which step two gets reversed in order to measure along the basis of the training vectors and retrieve the most likely (i.e. close) candidate. However, for classification problems we are fortunately not interested in the actual features of the nearest neighbours of $\ket{x_1...x_n}$, but merely in their class assignment. In superposition  $\ket{\psi_4}$, the different classes appear weighted by their member's distance to the input that is to classify. This is obvious if we rewrite state $\ket{\psi_4}$ as
\begin{multline*}\ket{\psi_4} = \frac{1}{\sqrt{N}} \sum \limits_{c=1}^d \ket{c} \otimes \sum \limits_{l\in c}  \mathrm{cos}\left[ \frac{\pi}{2n} \bar{d}_H(\vec{x},\vec{v}^l) \right]  \ket{x_1...x_n;d_1^{l} ...d_n^{l}; 0} \\ +  \mathrm{sin}\left[ \frac{\pi}{2n} \bar{d}_H(\vec{x},\vec{v}^l) \right]  \ket{x_1...x_n;d_1^{l} ...d_n^{l}; 1}  , \end{multline*}
where $l$  runs over all training vectors classified with the label $c$. The probability to measure a certain class $c \in \{1,...,d\}$ provided we previously measured the ancilla in $0$ is given by
\begin{equation}\mathrm{P}(c) = \frac{1}{N P(0)} \sum \limits_{l\in c} \mathrm{cos}^2\left[ \frac{\pi}{2n} \bar{d}_H(\vec{x},\vec{v}^l) \right], \label{result}\end{equation}
 a value that scales with the average Hamming distance between the input and all training vectors in this class. If we measure the class qudit of a sufficient number of copies of superposition $\ket{\psi_4}$, we can consequently retrieve the optimal class label for $\ket{x_1...x_n}$. This can be further processed as classical information, or as a new training vector $\ket{v_1^{P+1},...,v_n^{P+1}, c^{P+1}}= \ket{x_1...x_n, c_x}$ if we discard the qubits $d_1 ...d_n$ in the second register.\\

The second version would go as in \cite{trugenberger01,trugenberger02}, only that we are not interested in the closest training vector, but in the class of a number of close vectors. As in kNN, we assign the class that is the most represented amongst the neighbours. The difference to the classical algorithm is thereby that we do not necessarily pick the $k$ nearest neighbours, but any neighbours with a probability that is proportional to their proximity to the input vector.

\section{Discussion}
The quantum pattern classification algorithm sketched above  runs in polynomial time $\mathcal{O}(TPn)$\footnote{The complexity of a quantum algorithm is measured through the number of elementary gates that have to be applied to simulate the quantum evolution.} where $n$ is the size of the feature vectors, $P$ is the number of training examples and $T$ is an accuracy threshold. More precisely, we have $4n+2$ operations for the retrieval algorithm (the unitary $U$ can be decomposed into $2n$ elementary operations \cite{trugenberger02}), which we run $T$ times to get a sufficiently precise picture from the measurement results. The construction of the superposition lies in $\mathcal{O}(P n)$ \cite{ventura00,trugenberger02}. As a rough comparison, the classical kNN also has to compute the distance to all $P$ $n$-dimensional training examples, which leads to a similar complexity class. An interesting point is that if we find a more efficient way to construct the superposition $\ket{T}$ in $\mathcal{O}(n)$, or receive it from a quantum memory device, the quantum version of this pattern classification algorithm would be independent of the number of training vectors, something that seems impossible to achieve in a classical version. In addition to this, the distance weighting (assigning a weight to each neighbour) does not require an additional step, but is `combined' with the measuring of the distances.\\

\begin{figure}[t]
  \centering    \includegraphics[width=0.4\textwidth]{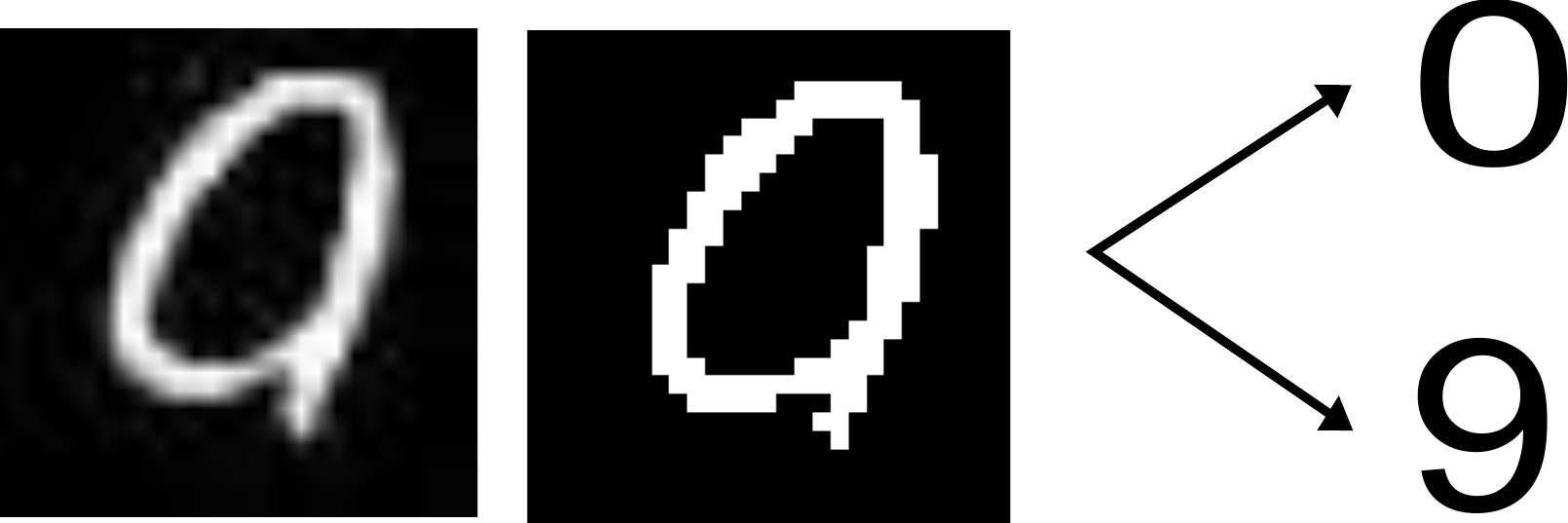} \caption{Example of an ambiguous input image for the task of handwritten digit recognition. The image is taken from the training set of the MNIST database and shows the original (left) and binarised (right) example of a handwritten `9' that can easily be recognised as a $0$ or a $9$.} \label{figure3}
\end{figure}

To illustrate another advantage of the quantum pattern classification algorithm, we consider the problem of recognising (in other words, classifying) handwritten digits, for example from the above mentioned MNIST handwritten digit database. Of course, running the algorithm on the unpreprocessed data of the binarised grey-shade pixel is only as successful as any classical algorithm executing a majority decision based on the Hamming distance between input and training vectors. This is without question a rather imprecise approach and our simulations show that approximately $50\%$ of the digits of a test set of $100$ examples can be classified correctly by this method (using a training set of 400 examples), a value that can be slightly improved by a scaling parameter $\epsilon$ introduced through a global phase shift in Eq (\ref{btw}). Still, the MNIST example helps to demonstrate how quantum computing offers a general advantage in cases of ambiguous inputs. Consider an image of a handwritten $9$ that is easily mistaken for a $0$, especially when applying a rough classification method based on the Hamming distance (see Figure \ref{figure3}). The classical kNN algorithm would lead to a deterministic output of either $0$ or $9$. On the other hand, repeating the quantum algorithm several times would lead to a distribution of outputs governed by Eq (\ref{result}), and we would expect the $0$ and $9$ to be almost equally frequent. As a consequence, the quantum algorithm produces additional information on the quality of the judgement in a classification task. In other words, the probabilistic output of quantum algorithms presents an asset for pattern classification, as can be shown through a method as simple as the one presented here. Future works would have to extend the quantum algorithm to allow for continuous inputs, and investigate ways to exploit its advantages using more complex distance measures. This is beyond the scope of this publication, in which we merely intend to demonstrate the potential of pattern classification through quantum information. In general, methods of quantum machine learning might become an important extension to the field of  machine learning, and create an exciting opportunity for both quantum physicists and computer scientists.

\section*{Acknowledgements}
This work is based upon research supported by the South African Research Chair Initiative of the Department of Science and Technology and National Research Foundation.


\begin{thebibliography}{10}

\bibitem{lloyd13}
Seth Lloyd, Masoud Mohseni, and Patrick Rebentrost.
\newblock Quantum algorithms for supervised and unsupervised machine learning.
\newblock {\em arXiv preprint arXiv:1307.0411}, 2013.

\bibitem{rebentrost13}
Patrick Rebentrost, Masoud Mohseni, and Seth Lloyd.
\newblock Quantum support vector machine for big feature and big data
  classification.
\newblock {\em arXiv preprint arXiv:1307.0471}, 2013.

\bibitem{wiebe14}
Nathan Wiebe, Ashish Kapoor, and Krysta Svore.
\newblock Quantum nearest-neighbor algorithms for machine learning.
\newblock {\em arXiv preprint arXiv:1401.2142}, 2014.

\bibitem{sentis12}
G~Sent{\'\i}s, J~Calsamiglia, Ram{\'o}n Mu{\~n}oz-Tapia, and E~Bagan.
\newblock Quantum learning without quantum memory.
\newblock {\em Scientific Reports}, 2, 2012.

\bibitem{gutja10}
M{\u{a}}d{\u{a}}lin Gu{\c{t}}{\u{a}} and Wojciech Kot{\l}owski.
\newblock Quantum learning: asymptotically optimal classification of qubit
  states.
\newblock {\em New Journal of Physics}, 12(12):123032, 2010.

\bibitem{rigatos07}
Gerasimos~G Rigatos and Spyros~G Tzafestas.
\newblock Neurodynamics and attractors in quantum associative memories.
\newblock {\em Integrated Computer-Aided Engineering}, 14(3):225--242, 2007.

\bibitem{busemeyer12}
Jerome~R Busemeyer and Peter~D Bruza.
\newblock {\em Quantum models of cognition and decision}.
\newblock Cambridge University Press, 2012.

\bibitem{sasaki01}
Masahide Sasaki, Alberto Carlini, and Richard Jozsa.
\newblock Quantum template matching.
\newblock {\em Physical Review A}, 64(2):022317, 2001.

\bibitem{dudani76}
Sahibsingh~A Dudani.
\newblock The distance-weighted k-nearest-neighbor rule.
\newblock {\em Systems, Man and Cybernetics, IEEE Transactions on},
  (4):325--327, 1976.

\bibitem{trugenberger01}
Carlo~A Trugenberger.
\newblock Probabilistic quantum memories.
\newblock {\em Physical Review Letters}, 87:067901, Jul 2001.

\bibitem{nielsen10}
Michael~A Nielsen and Isaac~L Chuang.
\newblock {\em Quantum computation and quantum information}.
\newblock Cambridge university press, 2010.

\bibitem{alpaydin04}
Ethem Alpaydin.
\newblock {\em Introduction to machine learning}.
\newblock MIT press, 2004.

\bibitem{hamming50}
Richard~W Hamming.
\newblock Error detecting and error correcting codes.
\newblock {\em Bell System Technical Journal}, 29(2):147--160, 1950.

\bibitem{harrington12}
Peter Harrington.
\newblock {\em Machine learning in action}.
\newblock Manning Publications Co., 2012.

\bibitem{rogers12}
Simon Rogers and Mark Girolami.
\newblock {\em A first course in machine learning}.
\newblock CRC Press, 2012.

\bibitem{duda12}
Richard~O Duda, Peter~E Hart, and David~G Stork.
\newblock {\em Pattern classification}.
\newblock John Wiley \& Sons, 2012.

\bibitem{hechenbichler04}
Klaus Hechenbichler and Klaus Schliep.
\newblock Weighted k-nearest-neighbor techniques and ordinal classification.
\newblock 2004.

\bibitem{giovannetti08}
Vittorio Giovannetti, Seth Lloyd, and Lorenzo Maccone.
\newblock Quantum random access memory.
\newblock {\em Physical Review Letters}, 100(16):160501, 2008.

\bibitem{schuld14c}
Maria Schuld, Ilya Sinayskiy, and Francesco Petruccione.
\newblock Quantum machine learning, 2014.
\newblock accepted for publication in Contemporary Physics.

\bibitem{lu14}
Songfeng Lu and Samuel~L Braunstein.
\newblock Quantum decision tree classifier.
\newblock {\em Quantum Information Processing}, 13(3):757--770, 2014.

\bibitem{lloyd13b}
Seth Lloyd, Masoud Mohseni, and Patrick Rebentrost.
\newblock Quantum principal component analysis.
\newblock {\em arXiv preprint arXiv:1307.0401v2}, 2013.

\bibitem{servedio04}
Rocco~A Servedio and Steven~J Gortler.
\newblock Equivalences and separations between quantum and classical
  learnability.
\newblock {\em SIAM Journal on Computing}, 33(5):1067--1092, 2004.

\bibitem{ventura00}
Dan Ventura and Tony Martinez.
\newblock Quantum associative memory.
\newblock {\em Information Sciences}, 124(1):273--296, 2000.

\bibitem{zhou08}
Rigui Zhou and Qiulin Ding.
\newblock Quantum pattern recognition with probability of 100\%.
\newblock {\em International Journal of Theoretical Physics}, 47(5):1278--1285,
  2008.

\bibitem{aimeur07}
Esma A{\"\i}meur, Gilles Brassard, and S{\'e}bastien Gambs.
\newblock Quantum clustering algorithms.
\newblock In {\em Proceedings of the 24th international conference on machine
  learning}, pages 1--8. ACM, 2007.

\bibitem{horn02}
David Horn and Assaf Gottlieb.
\newblock Algorithm for data clustering in pattern recognition problems based
  on quantum mechanics.
\newblock {\em Physical Review Letters}, 88(1):018702, 2002.

\bibitem{aimeur06}
Esma A{\"\i}meur, Gilles Brassard, and S{\'e}bastien Gambs.
\newblock Machine learning in a quantum world.
\newblock In {\em Advances in Artificial Intelligence}, pages 431--442.
  Springer, 2006.

\bibitem{gambs08}
S{\'e}bastien Gambs.
\newblock Quantum classification.
\newblock {\em arXiv preprint arXiv:0809.0444}, 2008.

\bibitem{neven09}
Hartmut Neven, Vasil~S Denchev, Geordie Rose, and William~G Macready.
\newblock Training a large scale classifier with the quantum adiabatic
  algorithm.
\newblock {\em arXiv preprint arXiv:0912.0779}, 2009.

\bibitem{schuetzhold02}
Ralf Sch{\"u}tzhold.
\newblock Pattern recognition on a quantum computer.
\newblock {\em arXiv preprint quant-ph/0208063}, 2002.

\bibitem{trugenberger02}
Carlo~A Trugenberger.
\newblock Quantum pattern recognition.
\newblock {\em Quantum Information Processing}, 1(6):471--493, 2002.

\bibitem{clark14}
Lewis~A Clark, Wei Huang, Thomas~M Barlow, and Almut Beige.
\newblock Hidden quantum markov models and open quantum systems with
  instantaneous feedback.
\newblock {\em arXiv preprint arXiv:1406.5847}, 2014.

\bibitem{barry14}
Jennifer Barry, Daniel~T Barry, and Scott Aaronson.
\newblock Quantum pomdps.
\newblock {\em arXiv preprint arXiv:1406.2858}, 2014.

\bibitem{sasaki02}
Masahide Sasaki and Alberto Carlini.
\newblock Quantum learning and universal quantum matching machine.
\newblock {\em Physical Review A}, 66(2):022303, 2002.

\bibitem{hunziker03}
Markus Hunziker, David~A Meyer, Jihun Park, James Pommersheim, and Mitch
  Rothstein.
\newblock The geometry of quantum learning.
\newblock {\em arXiv preprint quant-ph/0309059}, 2003.

\bibitem{bonner03}
Richard Bonner and Rusin{\v{s}} Freivalds.
\newblock A survey of quantum learning.
\newblock In {\em Quantum Computation and Learning}, page 106, 2003.

\end{thebibliography}
\end{document}